\documentclass[pmlr]{jmlr}

\usepackage{longtable}
\usepackage{mdframed}
\usepackage{booktabs}
\usepackage[load-configurations=version-1]{siunitx} 

\theorembodyfont{\upshape}
\theoremheaderfont{\scshape}
\theorempostheader{:}
\theoremsep{\newline}

\jmlrvolume{1}
\jmlryear{2024}
\jmlrworkshop{AAAI AI4Edu 24}

\title[Explaining Code Examples]{Explaining Code Examples in Introductory Programming Courses: LLM vs Humans}

   \author{
  \Name{Arun-Balajiee Lekshmi-Narayanan} 
  \thanks{Both authors contributed equally.}
  \Email{arl122@pitt.edu}\\
  \Name{Priti Oli}  \footnotemark[1]
  \Email{poli@memphis.edu}\\
  \Name{Jeevan Chapagain} \Email{jchpgain@memphis.edu}\\
  \Name{Mohammad Hassany} \Email{moh70@pitt.edu}\\
  \Name{Rabin Banjade} \Email{rbnjade1@memphis.edu}\\
  \Name{Peter Brusilovsky} \Email{peterb@pitt.edu}\\
  \Name{Vasile Rus} \Email{vrus@memphis.edu}\\
  \addr {University of Pittsburgh, PA, USA. 15260} \\
  \addr University of Memphis, TN, USA, 38152}

\begin{document}

\maketitle

\begin{abstract}
Worked examples, which present an explained code for solving typical programming problems are among the most popular types of learning content in programming classes. Most approaches and tools for presenting these examples to students are based on line-by-line explanations of the example code. However, instructors rarely have time to provide explanations for many examples typically used in a programming class. In this paper, we assess the feasibility of using LLMs to generate code explanations for passive and active example exploration systems. To achieve this goal, we compare the code explanations generated by chatGPT with the explanations generated by both experts and students.
\end{abstract}

\section{Introduction}
\label{sec:intro}

Program code examples (known also as worked examples) play a crucial role in learning how to program~\citep{linn1992CACM}.
Instructors use examples extensively to demonstrate the semantics of the programming language
being taught and to highlight the fundamental coding patterns. Programming textbooks allocate considerable space to present and explain code examples. To make the process of studying code examples more interactive, CS education researchers developed a range of tools to engage students in the study of code examples. These tools include codecasts~\citep{CODECAST2017}, interactive example explorers~\citep{Hosseini2020}, and tutoring systems~\citep{oli2023improving}. 

An important component in all types of program examples is code explanations associated with code lines or chunks. The explanations connect examples with general programming knowledge explaining the role and function of code fragments or their behavior.  In textbooks, these explanations are usually presented as comments in the code or as explanations on the margins. The example explorer tools allow students to examine these explanations interactively~\citep{Hosseini2020}. Tutoring systems, which engage students in explaining the code, use instructor explanations to assess student responses \citep{chapagain2022automated} and provide scaffolding~\citep{oli2023improving}. The explanations must be \emph{authored} by instructors or domain experts, i.e., written and integrated into a specific system. As the experience of the last 10 years demonstrated, these explanations are hard to obtain.
Being enthusiastic about sharing \emph{the code} of their worked examples with others, instructors generally do not have time or patience to properly author \emph{explanations} of their examples. Indeed, creating just one explained example could take 30 minutes even in the presence of authoring tools ~\citep{Hosseini2020,CODECAST2017}. As a result, the volume of worked examples available to students in a typical introductory programming class is low. 

To address this \emph{authoring bottleneck}, researchers explored \emph{learner-sourcing}, that is, engaging students in the creation and review of explanations of instructor-provided code~\citep{hsiao2011role} and the automatic extraction of explanations from lecture recordings~\citep{khandwala2018codemotion}. In this paper, we explore the feasibility of human-AI collaboration in creating explained code examples. With this approach, the instructor provides the code for their favorite examples. The AI engine based on large language models (LLM) examines the example code and generates explanations for each code line. The explanations are reviewed and, if necessary, edited by the instructor. 

To assess the feasibility of this approach, it is important to compare the code explanations produced by LLMs such as ChatGPT with explanations produced by humans. To use ChatGPT explanations in example explorer systems, we need to check how similar they are in language, semantics, and style used to code explanations produced by instructors and domain experts. To use ChatGPT explanations to assess student responses in tutoring systems, we need to assess how close these explanations are to the explanations produced by students when explaining the code. In this paper, we employ a range of analytical approaches to compare ChatGPT code explanations with explanations produced by both experts and students. Following a review of past work on using LLM for code explanations, we present the method and datasets used in the study and review the results. 

\section{Related Work: Use of LLMs for Code Explanations}
Multiple researchers have explored code summarization~\citep{Phillips2022ImprovedEO} and explanations using transformer models~\citep{10113620, peng-etal-2022-rethinking}, abstract syntax trees~\citep{Shi2022CodeDKTAC}, and Tree-LSTM~\citep{tian2023chatgpt}. With the announcement of ChatGPT, several research teams explored the use of LLM for code explanations using ChatGPT 3~\citep{10.1145/3544548.3581388,10.1145/3545945.3569785,leinonen2023comparing}, GPT 3.5~\citep{10.1145/3545945.3569785,li2023explaining,10.1007/978-3-031-36336-8_50}, GPT 4~\citep{li2023explaining}, OpenAI Codex \citep{10.1145/3501385.3543957,tian2023chatgpt,10.1145/3545945.3569785}, and GitHub Copilot \citep{10.1007/978-3-031-36336-8_50}. These LLMs were used to generate explanations at different levels of abstraction (line-by-line, step-by-step, and high-level summary). \cite{10.1145/3501385.3543957} observed that ChatGPT can generate better explanations at line-by-line level. 

The explanations and summaries generated by these LLMs were mostly evaluated by authors~\citep{10.1145/3501385.3543957}, students~\citep{10.1145/3545945.3569785,leinonen2023comparing}, and tool users~\citep{10.1007/978-3-031-36336-8_50}. 
Most recently, attempts have been made to compare code explanations generated by humans and LLMs. 
\cite{10.1145/3501385.3543957} reported that students rated LLM-generated explanations as useful, easier, and more accurate than learner-sourced explanations~\citep{leinonen2023comparing}. 
In this work, we attempted a more formal approach to compare the line-by-line code explanations generated by ChatGPT, students, and experts using a range of quantitative metrics. Our goal was to generate insights into the use of ChatGPT code explanations in the learning process.

\section{Dataset Collection}
To produce the evaluation dataset, we collected line-by-line explanations from three types of sources - experts, ChatGPT, and students - for four Java worked examples selected from different topics of the PCEX example exploration system~\citep{Hosseini2020}. The examples represent programming problems of different difficulty levels. The simpler program involved array search and print statements, while the hardest program focused more on object-oriented principles. The selected examples were used in multiple Java classes and included line-by-line explanations produced by the instructors. The summary of the types of explanation collected is shown in Table~\ref{tab:data_line_by_line_explanations} and the collection process is explained below. Figure~\ref{fig:a1-sample-data} shows sample explanations from different sources. 

\emph{Expert Explanations:} We used one set of expert explanations available at PCEX and collected the second set of line-by-line explanations from different experts.

\begin{figure}[h!]
\begin{mdframed}
\noindent \textbf{Program:} PointTester;
\textbf{Line number:} 12;	
\noindent \textbf{Line code:} \texttt{private int y;}

\noindent \textbf{Expert1:} Every object of the Point class will have its own y-coordinate. Therefore, we need to declare an instance variable for the class to store the y-coordinate of the point. We declare it as int because we want to have integer coordinates for the point. Note that an instance variable is a variable defined in a class, for which each instantiated object of the class has a separate copy, or instance. 

\noindent \textbf{Expert2:} The instance variables are declared as private to prevent direct access to them from outside the class. In this way, no unexpected modifications to a Point object's data are possible.  


\noindent \textbf{S:} This line declares a private integer variable named "y" to store the y coordinate of a point. 

\noindent \textbf{A:} This line defines a private instance variable 'y' of type int in the Point class. It contributes directly to the program's objective of storing the y-coordinate of the point.

\noindent \textbf{E:} This line declares a private instance variable 'y' of type int in the Point class to store the y-coordinate of a point. Declaring the y-coordinate variable is essential for keeping track of the point's position and contributes directly to the program's objective of storing the point's coordinates.

\noindent \textbf{Student1:} initialize a private value inside the point class with no value yet 

\noindent \textbf{Student2:} Declares the private int variable y. 

\noindent \textbf{Student3:} Creates a private int that can only be accessed by class Point called int y 

...

\noindent \textbf{Student59:} private variable used to store the value entered into the value of the y coordinate.
\end{mdframed}
    \caption{Sample Data illustrating the various entries under each column in our dataset.}
    \label{fig:a1-sample-data}

\end{figure}


\emph{Student Explanations:} We performed a user study in which students of a Java programming course were asked to write explanations for each line of the code examples selected for the study. In total, we collected line-by-line explanations from 60 students. For example, for the line of code \textit{``private int y''} a student participating in the study explained \textit{``Creates a new object class called Point''}. 


\emph{ChatGPT  Explanations:} We performed a sequence of internal studies to build ChatGPT prompts, which can produce the most useful line-by-line example explanations. For the final evaluation, we selected three prompts that produced good explanations on three different levels of detail. We used gpt-3.5-turbo to generate four sets of line-by-line explanations for each selected example using each of these prompts. To increase the diversity of explanations, the first set was generated with temperature value 0 and three additional sets were generated with temperature value 1 each time clearing the history.

\textbf{Simple Prompt:}~\label{lab:basic-prompt} The \emph{Simple prompt} used in the study included the code of the worked example and the following instruction: \textit{``Provide a line-by-line self-explanation for each line of code in the Java program above''}. The explanations generated by this prompt are concise and elicit the goal of each line of code quite well. 

\begin{figure}[h!]
    \begin{mdframed}
    \centering
    \includegraphics[width=0.65\textwidth]{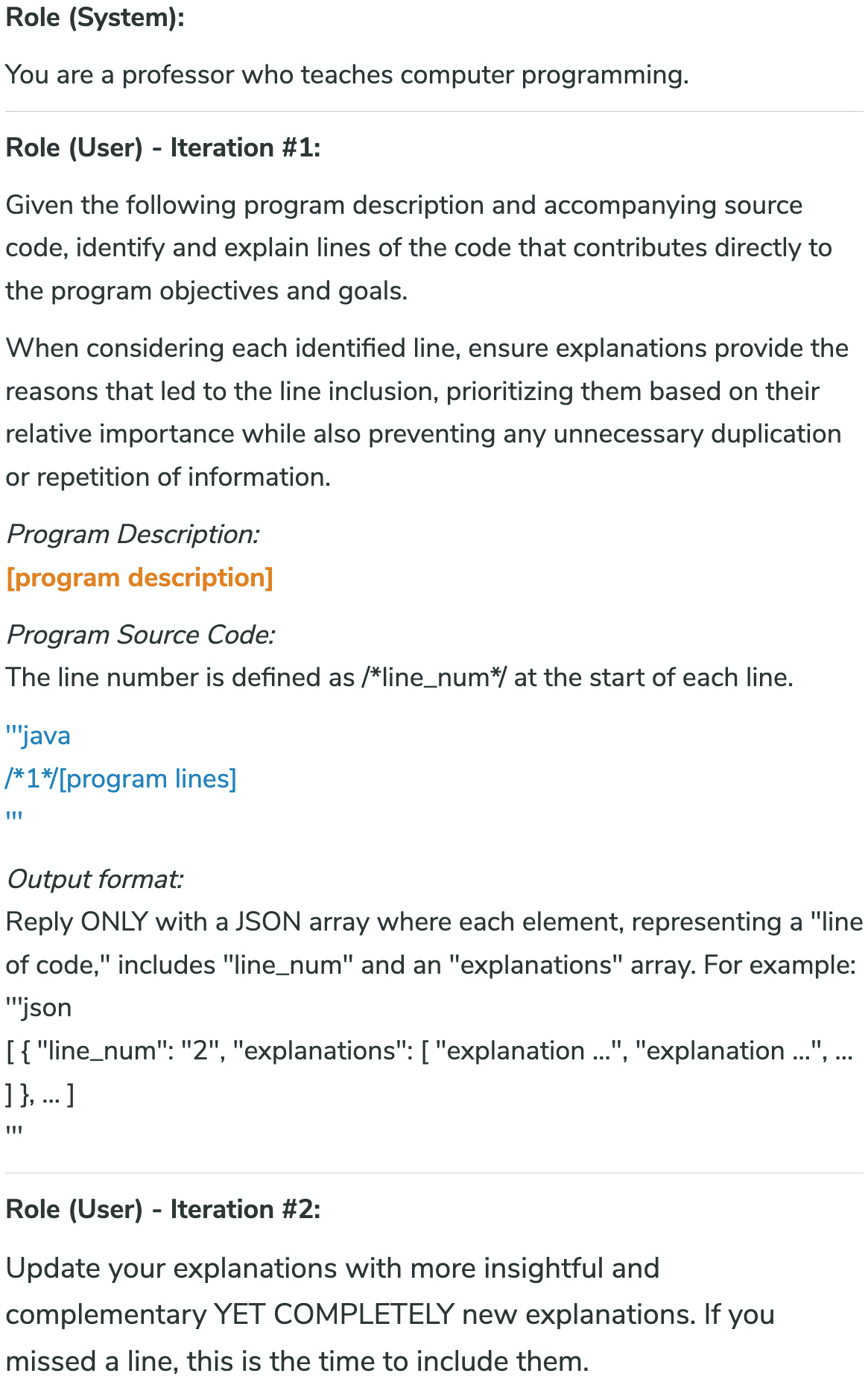}
    \end{mdframed}
    \caption{This ChatGPT Prompt template considers the case that ChatGPT could generate a better explanation with an additional ``nudge'' as observed above. In most cases, the generated explanations using the prompt at the second iteration produces richer explanation than the first iteration.}
    \label{fig:mohammad_prompt_template}
\end{figure}

\textbf{Advanced Prompt:} The \emph{Advanced prompt} used both the problem statement and the example code along with more elaborate instructions for ChatGPT. The role of ``a professor who teaches computer programming'' is assigned to the system to provide a context. 
The prompt also asked ChatGPT to provide reasons why the line needs to be explained. We observed that ChatGPT cannot always associate line numbers correctly, so each line in the program source code was annotated with its associated line number. An output format was defined to process the results digestible by our automation script. Prompt details are provided in Figure~\ref{fig:mohammad_prompt_template} (Iteration \#1 ).

\textbf{Extended Prompt:} 
To obtain the most elaborate ChatGPT explanations, we used \emph{Extended prompt}, which requested ChatGPT to further enhance the explanations generated by the Advanced prompt (Iteration \#2 in Figure~\ref{fig:mohammad_prompt_template}), with a focus on consistency and coverage of the generated content.


    %

\begin{table}[t!]
    \centering
    \begin{tabular}{l|l|l}\hline
         \textbf{Explanation Type} & \textbf{N} &\textbf{Definition} \\ \hline
         Experts & 2 & Source Code Line-by-Line Explanations by Experts \\ 
         Students & 60 &   Source Code Line-by-Line Explanations by Students \\ 
      Simple Prompt  (S) & 4 & ChatGPT Explanations with simple prompt (Section~\ref{lab:basic-prompt}) \\ 
      Advanced Prompt (A) & 4& ChatGPT Explanations with advanced prompt (Figure~\ref{fig:mohammad_prompt_template}) \\ 
      Extended Prompt (E) & 4&  ChatGPT Explanations with extended prompt (Figure~\ref{fig:mohammad_prompt_template}) \\ 
      ChatGPT & 12& Aggregated representation for all ChatGPT prompts \\ \hline
    \end{tabular} 
    \caption{A summary of explanation sources used in the study. 
    }
    \label{tab:data_line_by_line_explanations}
\end{table}

\section{Evaluation Metrics}

\emph{Lexical Metrics:} We report the \textit{lexical diversity} and \textit{lexical density} of the generated explanations to assess the richness, informativeness, and conciseness of the generated text ~\citep{johansson2008lexical}. \textit{Lexical diversity} is the range of
variety of distinct words or vocabulary used within a specific text. \textit{Lexical density} refers to the measure of the variety of different lexical words present in a text, including nouns, adjectives, verbs, and adverbs, which collectively contribute to the overall meaning of the text.  A recent paper considers the use of lexical diversity as a metric~\citep{Cegin2023ChatGPTTR} to compare content generated by humans and ChatGPT. We also report on the total number of tokens on each explanation to provide more insight into the comparison of the lexical features across each source. 

\emph{Readability Metrics:} We consider 3 popular metrics~\citep{errorMsgDenny1, errorMsgDenny2}, namely, Flesch-Kincaid Grade Level, Gunning Fog and Flesch Reading Ease. These metrics estimate the grade level expectation of the text such as the years of formal education required to read the text and the ease of reading a piece of text respectively. We use the TextDescriptives~\citep{Hansen2023TextDescriptivesAP} package in Python~\footnote{https://hlasse.github.io/TextDescriptives/readability.html} to calculate these scores.

\textit{Similarity Metrics:} We use four evaluation metrics: Character-based metric chrF~\citep{popovic2015chrf}, Word-based metric METEOR~\citep{banerjee2005meteor}, and Embedding-based metrics BERTScore~\citep{zhang2019bertscore} and Universal Sentence Encoder (USE)~\citep{cer2018universal} to compare explanation generated by different sources. chrF (character n-gram F-score) measures the character-level matching between the reference text and the machine-generated text considering both precision and recall. METEOR considers the similarity between words and assesses word overlap between the two texts.
BERTScore is an automated evaluation metric for text generation that assesses the similarity between candidate and reference sentences by comparing the contextual embeddings of individual tokens using cosine similarity.
USE is a transformer-based model that transforms text into high-dimensional vectors, enabling the computation of similarity between two texts based on their vector representations.
\cite{haque2022semantic} and \cite{roy2021reassessing} have pointed out that METEOR, chrF~\citep{popovic2015chrf}, and USE~\citep{cer2018universal} metrics are better aligned with human preferences of code summarization as these metrics assign partial credits to words. We also use BertScore to evaluate the generated explanation primarily due to its extensive use as a reliable measure for evaluating the faithfulness of LLMs~\citep{ji2023survey}. Consequently, traditional metrics like BLEU~\citep{papineni2002bleu} which solely rely on word overlap, are now considered outdated and are not included in our reporting.




\section{Results}



    



We compared the explanations using metrics aggregated over all 33 explainable lines of four examples generated by each expert, each student, and each round of ChatGPT generation  Table~\ref{tab:lexcial_metric_detail} reports the medians for each source \emph{type}. 

\emph{Lexical Metrics:} 
As Table~\ref{tab:lexcial_metric_detail} shows, explanations produced by experts and ChatGPT are more than twice as long as explanations produced by students (as measured by the number of tokens). Students also use considerably fewer unique words in their explanations (lexical diversity) hinting that their vocabulary is more narrow than the vocabulary of experts and ChatGPT. The length and lexical diversity of explanations generated by ChatGPT and experts varied, with \emph{Simple prompt} generating the shortest, \emph{Extended prompt} the longest explanations, and expert explanations positioned between \emph{Advanced} and \emph{Extended} prompts.
An ANOVA analysis of the lexical diversity of the explanation generated by experts, ChatGPT, and students indicated statistically significant variations among the groups (F-statistic = 25.07, $p < 0.05$). 
The data also show that the explanations produced by the students are not only shorter than those by experts and ChatGPT, but also have a higher lexical density, suggesting that the students explain the code in a more ``concentrated'' way. However, the ANOVA analysis of the lexical density does not indicate any significant difference (F-statistic = 2.5, p =0.08) between the explanations in terms of lexical density. 


\begin{table}[h!]
    
    \begin{tabular}{c c c c c c c c }
    \hline
         Source &  N & Vocabulary & Lexical Density  & \# of Tokens & GF & FRE & FK\\
         \hline
            Experts & 2 & 209.0  & 0.48 & 690.0 & 8.46 & 78.45 & 6.18\\
            S & 4 & 165.0 & 0.45 & 517.5 & 8.67 & 82.34 & 6.35	\\
            A & 4 & 185.5 & 0.48 & 625.0 & 9.91 & 72.63 & 7.15	\\
            E & 4 & 238.0  & 0.49 & 769.5 & 11.09 & 69.64 & 7.83	\\
            ChatGPT & 12 &  179.5 & 0.48 & 625.0 & 8.99 & 75.41 & 6.69	\\
            Students & 60 & 116.5 & 0.54 & 249.5 & 8.02 & 80.48 & 5.62 \\
\hline
\end{tabular}
    \caption{Median lexical and readability metrics for different sources of explanations (FRE = Flesch-Reading Ease, FK = Flesch-Kincaid, GF = Gunning Fog)} 
    \label{tab:lexcial_metric_detail}
\end{table}


\emph{Readability Metrics:} One-way ANOVA revealed that the Gunning-Fog readability scores are significantly different between explanation sources (experts, students, and ChatGPT) ($p < 0.001$). From post hoc comparisons, the most significant differences were observed between explanations produced by \emph{Extended prompt} and experts ($p  = 0.0102$) as well as students ($p < 0.0001$). No significance was observed between experts and students ($p = 0.0848$). Overall, for all metrics, the ChatGPT explanations are relatively less readable
than those of experts, which are less readable than those of students (Table~\ref{tab:lexcial_metric_detail}). 



\emph{Similarity Metrics:} We applied similarity metrics to calculate the similarity between the explanations provided by ChatGPT, students and experts for each line of code. We computed pairwise similarity scores between ChatGPT, Expert, and Student explanations averaged over all the 33 lines from all 4 programs. Table~\ref{tab:similarity_metric_table} shows that the explanations generated by ChatGPT exhibit a consistently higher average similarity score to the expert explanations compared to those generated by students. This suggests that the explanations generated by ChatGPT are more closely aligned with expert explanations than with student explanations across all metrics. 



Mann-Whitney U-tests indicate  significantly higher alignment between ChatGPT and expert explanations than between student and expert explanations (F-statistic = 48.0, $p<0.05$ for METEOR, F-statistic = 205.0, $p<0.05$ for USE and F-statistics=288.0, $p<0.05$ for BERTScore). We also observed that \emph{Simple} prompt (S) generated explanations aligned more closely with expert explanations than \emph{Advanced} and \emph{Extended} prompts (A and E). 
The higher semantic alignment between the expert explanation and the ChatGPT explanations generated with a \emph{Simple} rather than \emph{Advanced} prompt could be explained by the nature of the \emph{Advanced} prompt, which was specifically engineered to produce very detailed explanations which experts rarely have time to produce. In particular, explanations generated using \emph{Advanced} prompt frequently explained why each line of the code is important, while neither expert explanations nor explanations generated with \emph{Simple} prompt consistently explained the importance of each line in the code. 








\begin{table}[h!]
    \centering
    \begin{tabular}{ c c c c c c}
    \hline
        Reference & Source & chrF & METEOR & USE & BERTScore \\
    \hline
        Expert & ChatGPT(S)  & 0.32 & 0.28 & 0.54 & 0.89 \\
        Expert & ChatGPT(A)  & 0.31 & 0.27 & 0.49 & 0.709 \\
        Expert & ChatGPT(E)  & 0.32 & 0.28 & 0.48 & 0.712 \\
        Expert & ChatGPT (All)  & 0.32 & 0.27  & 0.50 & 0.75 \\
        Expert & Student  & 0.33 & 0.144 & 0.33 & 0.63 \\
\midrule
        ChatGPT(S) & Student & 0.22 & 0.21 & 0.34 & 0.60 \\
        ChatGPT(A) & Student & 0.18 & 0.150 & 0.254 & 0.450 \\
        ChatGPT(E) & Student & 0.18 & 0.151 & 0.255 & 0.458 \\
        ChatGPT(All) & Student & 0.19 & 0.17  & 0.28 & 0.50 \\
\midrule
        ChatGPT(S) & ChatGPT(A)  & 0.33 & 0.30 & 0.50 & 0.72 \\
        ChatGPT(S) & ChatGPT(E)  & 0.32 & 0.28 & 0.50 & 0.73 \\
        ChatGPT(A) & ChatGPT(E) & 0.44 & 0.43 & 0.56 & 0.69 \\
    
    \hline
    \end{tabular}
    \caption{Assessing alignment (larger is better) between sources of explanations}
    \label{tab:similarity_metric_table}
\end{table}



\section{Conclusions and Future Work}

In this work, our goal was to assess the feasibility of using ChatGPT to generate line-by-line code explanations to be used in worked-out examples in place of currently used expert explanations. To achieve this goal, we compared the ChatGPT explanations generated by different prompts with the student and expert explanations using a range of metrics. Our results indicate that the explanations generated by ChatGPT are lexically and semantically similar to the explanations generated by experts and could potentially resolve the authoring bottleneck. However, their lower readability level might be an obstacle for less-prepared students.  We also observed a considerable difference between the explanations produced by students and explanations produced by experts and ChatGPT, which might affect the efficiency of both sources of explanation in active example tutors where student explanations are assessed by comparing them with expert explanations. Both issues require a deeper investigation, which we plan to perform through a user study.

\section{Acknowledgement}
This  work  has  been  supported  by NSF awards 1822816 and 1822752. The opinions, findings, and results are solely those of the authors and do not reflect those of NSF.

\bibliography{references}


\appendix

\end{document}